\documentclass[notitlepage]{article}
\usepackage{amssymb}
\usepackage{amsfonts}
\usepackage{amsmath}
\usepackage{eurosym}
\usepackage{setspace}
\usepackage{graphicx}

\newtheorem{theorem}{Theorem}

\newtheorem{claim}{Claim}

\newtheorem{condition}{Condition}

\newtheorem{corollary}{Corollary}

\newtheorem{lemma}{Lemma}

\newtheorem{remark}{Remark}

\input{tcilatex}
\begin{document}

\title{Aggregating Incomplete Rankings\thanks{%
The author is grateful to Minoru Kitahara, Masashi Umezawa, Susumu Cato, and
Takako Fujiwara-Greve for insightful comments and suggestions. This work was
supported by JSPS KAKENHI Grant Numbers 20K01675 and 22K01402.}}
\author{Yasunori Okumura\thanks{%
Address: 2-1-6, Etchujima, Koto-ku, Tokyo, 135-8533 Japan.
Phone:+81-3-5245-7300. Fax:+81-3-5245-7300. E-mail:
okuyasu@gs.econ.keio.ac.jp}}
\maketitle

\begin{center}
\textbf{Abstract}
\end{center}

This study considers the method to derive a ranking of alternatives by
aggregating the rankings submitted by several individuals who may not
evaluate all of them. The collection of subsets of alternatives that
individuals (can) evaluate is referred to as an evaluability profile. For a
given evaluability profile, we define an aggregating ranking function whose
inputs are the rankings of individuals on the alternatives they evaluate. We
investigate the properties of aggregating ranking functions, which are
modifications of those introduced in previous studies. Whether an
aggregating ranking function satisfying a combination of properties exists
depends on the evaluability profile. Thus, we identify the necessary and
sufficient conditions on evaluability profiles to ensure the existence of
the functions satisfying four different combinations of properties.
Furthermore, to examine how frequently possible or impossible evaluability
profiles occur, we derive the proportion of each type in specific cases.



\newpage

\section{Introduction}

This study examines situations in which the ranking of finite alternatives
is required, involving the rankings of these alternatives by a number of
individuals. Examples illustrating this situation include a faculty
considering several candidates for new members, or a company evaluating
various project proposals. In such organizations, decisions are often
influenced by the evaluation rankings submitted by their members. Thus, the
planner asks individuals to submit their rankings of alternatives, but the
rankings submitted by some individuals may be incomplete. In a real-world
example, the evaluation of a specific competitive research grant is
conducted through peer review. During the review process, evaluations of
researchers with whom there is a conflict of interest are designed to be
avoided through self-disclosure.

The reasons why some individuals do not or should not submit complete
ranking are such as those as follows. First, they may lack sufficient
information or knowledge to evaluate these alternatives. For example,
evaluating potential faculty members may require expertise in their specific
field. Second, an individual is known to be biased in favor of their family
members, colleagues, themselves, their own ideas, and the projects they were
previously involved in.\footnote{%
In psychology, such a bias is called \textquotedblleft ingroup
bias.\textquotedblright\ Similar biases are also frequently observed in the
behavioral economics literature. For instance, several studies are
introduced in Ariely's (2010) third and fourth chapters.
\par
{}Moreover, there are several voting models that consider the biases of
voters. For example, Amor\'{o}s et al. (2002), Amor\'{o}s (2009, 2011, 2019)
and Adachi (2014) consider a jury problem where some juries have some biases.%
} As a result, they may not be well-suited to provide objective evaluations
of these alternatives. Third, planners may encounter difficulties in
obtaining complete rankings from all individuals. For example, when new
alternatives are added after some individuals have already submitted their
rankings, there may be no opportunity to gather their rankings for the newly
added ones. Consequently, the evaluations by these individuals may not be
incorporated into the aggregated ranking of the alternatives.

To address such situations, we revisit a model originally introduced by Ohbo
et al. (2005) and also discussed by Ando et al. (2007). In the model, each
individual is assigned a subset of alternatives to evaluate and is asked to
submit a weak order (ranking) of them. In this study, we refer to the
collection of these subsets as an evaluability profile.\footnote{%
Several previous studies, including Quesada (2005) and Cengelci and Sanver
(2007) examine models in which voters have the option to abstain from
voting. In contrast, this study assumes that the evaluability profile is
independent of individuals' choices and is exogenously given.} Given an
evaluability profile, we define an aggregating (incomplete) ranking function
that takes the rankings submitted by all individuals as inputs. This
function is equivalent to the social welfare function \`{a} la Arrow (1963)
when the ranking of each individual is complete. Moreover, in the peer
rating model introduced by Ando et al. (2003) and Ng and Sun (2003),
individuals evaluate each other, but no individual evaluates themselves.%
\footnote{%
The prize award problem, as discussed by Holzman and Moulin (2013), Tamura
and Ohseto (2014), Mackenzie (2015, 2019), Tamura (2016) and Edelman and Por
(2021) is related to the peer rating problem.} Our model is also a
generalization of the peer rating model.

We examine six properties of aggregating ranking functions: transitive
valuedness, the Pareto criterion, the weak Pareto criterion, independence of
irrelevant alternatives, non-dictatorship, and non-consistency which are
modifications of those of Arrow (1963) and Hansson (1973).

First, we derive the necessary and sufficient condition on the evaluability
profile to ensure the existence of an aggregating ranking function that
satisfies transitive valuedness and the Pareto criterion. Second, we
additionally require independence of irrelevant alternatives but show that
the necessary and sufficient condition is unchanged. Third, we derive the
necessary and sufficient condition ensuring the existence of an aggregating
ranking function that satisfies non-dictatorship in addition to the
properties of the second result. We need an additional condition for the
existence of an aggregating ranking function that additionally satisfies
non-dictatorship. Fourth, we revisit the properties in the second result but
the weak Pareto criterion and non-consistency are considered instead of the
Pareto criterion, because the Pareto criterion implies them. However, the
necessary and sufficient condition is the same as that of the second result
as well as the first result.

The first result introduced above is a generalization of that of Ando et al.
(2003) which specifically focus on the peer rating model. Moreover, the
fourth result introduced above is also closely related to one of the results
of Ando et al. (2003). The proofs of these results are inspired by their
proof techniques. Furthermore, the third result is a generalization of that
of Ohbo et al. (2005) that consider the case involving at least one
individual who evaluates all alternatives. While the results of Ando et al.
(2003) and Ohbo et al. (2005) show impossibility only, we introduce specific
ARFs to have the conditions that are sufficient for the existence of ARFs.

By the second and third results, we can distinguish evaluability profiles
into three types. The first type is characterized by the absence of ARFs
that satisfy transitive valuedness, the Pareto criterion, and independence
of irrelevant alternatives. The second type is such that an ARF satisfying
the three must be a dictatorship rule. The third type is one for which there
is an ARF that satisfies non-dictatorship in addition to the three
properties. To examine how frequently each type occurs, we derive the
proportion of each type in specific cases.

There are three recent related studies to ours. First, Fujiwara-Greve et al.
(2023) independently discuss a model similar to ours. In their model,
described in our terminology, an evaluability profile is also an input of
the aggregating ranking function. They consider several combinations of
properties on the function, which differ from ours, and discuss which
combinations are possible or impossible for the existence of a function
satisfying them. On the other hand, in our model, for a given evaluability
profile, the aggregating ranking function is defined, and we consider the
conditions on the evaluability profiles.\footnote{%
Since in the combinations of axioms in this study whether impossible or
possible is dependent on the evaluability profile, these combinations are
\textquotedblleft impossible\textquotedblright\ in the sense of
Fujiwara-Greve et al. (2023).} Moreover, Fujiwara-Greve et al. (2023)
restrict their attention to a class of evaluability profiles. In this study,
we also clarify the results when we focus on the class that they consider.
Second, Barber\`{a} and Bossert (2023) introduce a general model including
the case where the rankings submitted by individuals may be incomplete. In
their approach, they aggregate the opinions of individuals without
identifying who submitted them. On the other hand, since we require
non-dictatorship in our analysis, we need to identify whose opinions they
are. Villar (2023) also introduces a similar general model that includes the
case of incomplete preferences and introduces aggregating functions that
provide a quantitative assessment to each alternative and satisfy a
combination of properties. On the other hand, our aggregating functions
provide just a weak order of alternatives, and thus their properties also
differ from ours.

\section{Model and technical results}

\subsection{Problems}

We consider a model that is originally introduced by Ohbo et al. (2005). Let 
$A$ be the finite set of alternatives. Let $V$ be the finite set of
individuals (voters) who submit a ranking of alternatives. We assume $%
\left\vert A\right\vert \geq 3$ and $\left\vert V\right\vert \geq 3$.

We consider the situation where some individuals may not evaluate some of
the alternatives. Let $A_{v}\subseteq A$ be an \textbf{evaluable set} of
individual $v$ representing the set of alternatives that $v$ (can) evaluate
and $\mathbf{A}=\left( A_{v}\right) _{v\in V}\subseteq A^{\left\vert
V\right\vert }$ be an \textbf{evaluability profile}. We assume that for any $%
v\in V$, at least two alternatives are in $A_{v}$. For $a\in A,$ let%
\begin{equation*}
V\left( a\right) =\left\{ v\in V\text{ }\left\vert \text{ }a\in A_{v}\right.
\right\} \text{,}
\end{equation*}%
which is the set of individuals who (can) evaluate\textit{\ }$a$. If $%
A_{v}=A $ (or equivalently if $v\in V\left( a\right) $ for all $a\in A$),
then individual $v$ is said to be \textbf{complete}.

An evaluability\textbf{\ }profile $\mathbf{A}=\left( A_{v}\right) _{v\in V}$
is said to be \textbf{larger} (resp. \textbf{smaller}) than another $\mathbf{%
A}^{\prime }=\left( A_{v}^{\prime }\right) _{v\in V}$ if $A_{v}^{\prime
}\subseteq A_{v}$ (resp. $A_{v}\subseteq A_{v}^{\prime }$) for all $v\in V$
and $A_{v}^{\prime }\subsetneq A_{v}$ (resp. $A_{v}\subsetneq A_{v}^{\prime
} $) for some $v\in V$.

We say that a \textbf{problem} is an ordered tuplet of the sets of
individuals and alternatives, and an evaluability profile $(V,A,\mathbf{A})$%
, which are the primitives of this study. Hereafter, we fix a problem $(V,A,%
\mathbf{A})$.

If $A_{v}=A$ for all $v\in V$, then the problem is equivalent to that of the
Arrow's (1963) classical problem. Moreover, suppose that $A=V;$ that is, the
sets of individuals and alternatives are equivalent, and $\mathbf{A}$ is
such that $A\setminus A_{v}=\left\{ v\right\} $ for all $v\in V$. Then, the
problem is equivalent to the peer rating problem, introduced by Ando et al.
(2003) and Ng and Sun (2003); that is, which considers the situation where
the individuals evaluate each other but each individual does not evaluate
themselves.

Fujiwara-Greve et al. (2023) call an evaluability\textbf{\ }profile $\mathbf{%
A}$ is \textbf{non-trivial} if for any pair of alternatives $a,b\in A$,
there is at least one $v\in V$ such that $a,b\in A_{v}$; that is, there is
no pair of alternatives that are evaluated by nobody. They focus only on the
non-trivial evaluability profiles.

\subsection{Binary relations}

For $A^{\prime }\subseteq A$, $R\subseteq A^{\prime }\times A^{\prime }$ be
a binary relation on $A^{\prime }$. If $\left( a,b\right) \in R$, then we
write $aRb$. Let $P\left( R\right) \,$and $I\left( R\right) \,$be the
asymmetric part and the symmetric part of a binary relation on $A^{\prime }$
denoted by $R;$ that is, 
\begin{eqnarray*}
P\left( R\right) &=&\left\{ \left. \left( a,b\right) \in A^{\prime }\times
A^{\prime }\text{ }\right\vert \text{ }aRb\text{ and }\lnot \left(
bRa\right) \right\} , \\
I\left( R\right) &=&\left\{ \left. \left( a,b\right) \in A^{\prime }\times
A^{\prime }\text{ }\right\vert \text{ }aRb\text{ and }bRa\right\} .
\end{eqnarray*}

A binary relation $R$ on $A^{\prime }\subseteq A$ is

\begin{description}
\item \textbf{reflexive}\textit{\ }if for all $a\in A^{\prime }$, $aRa$,

\item \textbf{complete}\textit{\ }if for all $a,b\in A^{\prime },$\textit{\ }%
$aRb$ or $bRa$ (or both),

\item \textbf{transitive} if for all $a,b,c\in A^{\prime },$ [$aRb$ and $bRc$%
] implies $aRc$,

\item \textbf{anti-symmetric }if for all $a,b\in A^{\prime },$ $aI\left(
R\right) b$ implies $a=b$,

\item \textbf{asymmetric} if for all $a,b\in A^{\prime },$ $aRb$ implies $%
\lnot \left( bRa\right) $,

\item \textbf{acyclic} if, for all $M\in \mathbb{N}\setminus \{1\}$ and all $%
a_{1},a_{2},\cdots ,a_{M}\in A^{\prime }$, $a_{m+1}Ra_{m}$ for all $m\in
\left\{ 1,\cdots ,M-1\right\} $ imply $\lnot \left( a_{1}Ra_{M}\right) $.
\end{description}

See, for example, Duggan (1999) for a detailed explanations on them. Note
that if $R$ (on $A^{\prime }$) is acyclic, then it is asymmetric.

A binary relation $R$ on $A^{\prime }$ is said to be a \textbf{weak order}
on $A^{\prime }$ if it is reflexive, complete and transitive. A binary
relation $R$ on $A^{\prime }$ is said to be a \textbf{linear order} on $%
A^{\prime }$ if it is reflexive, complete, transitive and anti-symmetric. If 
$R$ is a linear order on $A^{\prime }$, then we write 
\begin{equation*}
R:a_{1},a_{2},\cdots ,a_{\left\vert A^{\prime }\right\vert }
\end{equation*}%
representing $a_{x}Ra_{x+1}$ for all $x=1,\cdots ,\left\vert A^{\prime
}\right\vert -1$.

A binary relation $R^{\prime }$ on $A^{\prime }\subseteq A$ is a \textbf{%
linear order extension} of an asymmetric binary relation $R$ on $A^{\prime }$
if $R^{\prime }$ is a linear order on $A^{\prime }$ and $R\subseteq
R^{\prime }$. Since $R$ is asymmetric, a linear order extension of $R$ is
well-defined. Let $\mathcal{E}\left( R\right) $ be the set of all linear
order extensions of $R$ on $A^{\prime }$. The following fact is also known.
See, for example, Duggan (1999) and Bossert and Suzumura (2010).

\begin{remark}
If $R$ on $A^{\prime }\subseteq A$ is acyclic, then $\mathcal{E}\left(
R\right) \neq \emptyset $.
\end{remark}

Finally, for a binary relation on $A^{\prime }\subseteq A$ denoted by $R$
and $A^{\prime \prime }\subseteq A^{\prime }$, let 
\begin{equation*}
\left. R\right\vert _{A^{\prime \prime }}=R\cap \left( A^{\prime \prime
}\times A^{\prime \prime }\right) ,
\end{equation*}%
which is a binary relation on $A^{\prime \prime }$.

\subsection{Aggregating ranking functions and properties}

Let $\mathcal{W}_{v}$ be the set of all possible weak orders on $A_{v}$. Let 
$R_{v}\in \mathcal{W}_{v}$ represent a \textbf{ranking} on $A_{v}$ submitted
by $v$.\footnote{%
A voter $i$ may have an evaluation order $R_{v}^{\prime }$ on $A^{\prime
}\supsetneq A_{v},$ but he/she reveals a preference order on $A_{v}$. In
that case, $R_{v}$ $=\left. R_{v}^{\prime }\right\vert _{A_{v}}$.} Moreover,
let 
\begin{equation*}
\mathcal{W\equiv }\prod\limits_{v\in V}\mathcal{W}_{v}\text{ and }\mathbf{R}%
\left( \mathcal{\equiv }\left( R_{v}\right) _{v\in V}\right) ,\mathbf{R}%
^{\prime }\left( \mathcal{\equiv }\left( R_{v}^{\prime }\right) _{v\in
V}\right) \in \mathcal{W}.
\end{equation*}

Let $f:\mathcal{W}\rightarrow \mathcal{B}$ be an \textbf{aggregating ranking
function (hereafter ARF)}, where $\mathcal{B}$ is the set of all reflexive
and complete binary relations on $A$.

We introduce the properties of $f$ that are modifications of those
introduced by Arrow (1963). First, an ARF $f$ satisfies \textbf{transitive
valuedness }(hereafter \textbf{TV})\textbf{\ }if $f\left( \mathbf{R}\right) $
is transitive\textbf{\ }for all\textbf{\ }$\mathbf{R}\in \mathcal{W}$.

Second, we say that an ARF $f$ satisfies the \textbf{Pareto criterion }%
(hereafter \textbf{PC}), if for all $\mathbf{R}\in \mathcal{W}$, and all $%
a,b\in A,$ if $aP\left( R_{v}\right) b$ for all $v\in V\left( a\right) \cap
V\left( b\right) $, then $aP\left( f\left( \mathbf{R}\right) \right) b$.

Third, we say that $f$ satisfies the \textbf{weak} \textbf{Pareto criterion }%
(hereafter \textbf{wPC}), if for all $\mathbf{R}\in \mathcal{W}$, and all $%
a,b\in A,$ if $aP\left( R_{v}\right) b$ for all $v\in V\left( a\right) \cap
V\left( b\right) $, then $af\left( \mathbf{R}\right) b$ (but $bf\left( 
\mathbf{R}\right) a$ is allowed to hold). Ando et al. (2003) and Ohseto
(2007) consider this in the peer rating model.

Fourth, we say that $f$ satisfies \textbf{independence of irrelevant
alternatives }(hereafter \textbf{IIA}), if for all $a,b\in A$ and all $%
\mathbf{R,R}^{\prime }\in \mathcal{W},$ $\left. R_{v}\right\vert
_{\{a,b\}}=\left. R_{v}^{\prime }\right\vert _{\{a,b\}}$ for all $v\in
V\left( a\right) \cap V\left( b\right) $ implies $\left. f\left( \mathbf{R}%
\right) \right\vert _{\{a,b\}}=\left. f\left( \mathbf{R}^{\prime }\right)
\right\vert _{\{a,b\}}$. Note that for $a,b\in A$ such that $V\left(
a\right) \cap V\left( b\right) =\emptyset $, $\left. f\left( \mathbf{R}%
\right) \right\vert _{\{a,b\}}$ is dependent on irrelevant alternatives,
because in this case, there is no individual who evaluate both $a$ and $b$.

Fifth, we define a \textquotedblleft dictator\textquotedblright\ of this
model. First, individual $v$ is said to be a \textbf{quasi-dictator} of $f$
if for all $\mathbf{R}\in \mathcal{W}$ and\textbf{\ }all $a,b\in A,$ $%
aP\left( R_{v}\right) b$ implies $aP\left( f\left( \mathbf{R}\right) \right)
b$. Second, individual $v$ is said to be a \textbf{dictator} of $f$ if $v$
is a quasi-dictator of $f$ and complete. If $v$ is not complete, then the
positions of the alternatives in $A\setminus A_{v}$ are determined without
any opinion of $v$. Thus, in that case, $v$ is not called a dictator of $f$,
even if the rankings revealed by $v$ is always reflected in $f$. That is, in
this study, a quasi-dictator is a dictator only if they evaluates all
alternatives. We say that $f$ satisfies \textbf{non-dictatorship }(hereafter 
\textbf{ND}) if there is no dictator of $f$. Note that if there is no
complete individual , then any ARF satisfies\textbf{\ ND}, because nobody
can be a dictator for that profile in our definition.

Finally, we say that an ARF $f$ satisfies \textbf{non-constancy }(hereafter 
\textbf{NC}), if there is no pair of alternatives $\left( a,b\right) $ with $%
a\neq b$ and $V\left( a\right) \cap V\left( b\right) \neq \emptyset $ such
that $aP\left( f\left( \mathbf{R}\right) \right) b$ for all $\mathbf{R}\in 
\mathcal{W}$ (or $bP\left( f\left( \mathbf{R}\right) \right) a$ for all $%
\mathbf{R}\in \mathcal{W}$) or $aI\left( f\left( \mathbf{R}\right) \right) b$
for all $\mathbf{R}\in \mathcal{W}$. This property is a modification of that
introduced by Hansson (1973). Ando et al. (2003) also consider this property
in the peer rating model.

\begin{remark}
If an ARF $f$ satisfies \textbf{PC}, then it also satisfies \textbf{wPC }and 
\textbf{NC. }
\end{remark}

By this remark, when \textbf{PC }is required, we do not explicitly require 
\textbf{wPC }and \textbf{NC}. However, \textbf{wPC} does not imply \textbf{%
NC, }and vice versa.

Note that except for \textbf{wPC}, \textbf{ND},\textbf{\ }and\textbf{\ NC},
the definitions of the properties are the same as those introduced by Ohbo
et al. (2005). They do not consider \textbf{wPC }or\textbf{\ NC} and do not
explicitly define\textbf{\ ND}. However, since they show the existence of a
dictator, who is a quasi-dictator and a complete individual, they
(implicitly) focus on \textbf{ND}\ in our definition.

\subsection{Graphs}

Since we use some graph theory descriptions to introduce our results, we
first present some notations of graph theory. Throughout of this paper, we
let the set of alternatives $A$ be the set of nodes. An (undirected)\textbf{%
\ graph} is a set of unordered pairs of nodes. If $\left\{ a,b\right\} $ is
an element of a graph $G$, then we simply write $ab\in G$ or equivalently $%
ba\in G$. If $ab\in G$ for all $a,b\in A$ such that $a\neq b$, then $G$ is
said to be the \textbf{complete graph}.

A subset of graph $G$ denoted by $P\subseteq G$ is said to be an
(undirected) \textbf{path} of $G$ if 
\begin{equation*}
P=\left\{ a_{1}a_{2},a_{2}a_{3},\cdots ,a_{M-1}a_{M}\right\} ,
\end{equation*}%
where $a_{1},a_{2},\cdots ,a_{M}$ are distinct. A subset of graph $G$
denoted by $C\subseteq G$ is said to be an (undirected) \textbf{cycle} of $G$
if 
\begin{equation}
C=\left\{ a_{1}a_{2},a_{2}a_{3},\cdots ,a_{M-1}a_{M},a_{M}a_{1}\right\} ,
\label{c}
\end{equation}%
where $a_{1},a_{2},\cdots ,a_{M}$ are distinct and $M\geq 3$, which rules
out that $C=\left\{ a_{1}a_{2},a_{2}a_{1}\right\} $ is a cycle. For
notational convenience, when we use a cycle denoted by (\ref{c}), $%
a_{M+1}=a_{1}$. If all nodes are included in a cycle; that is, if $\left\{
a_{1},a_{2},\cdots ,a_{M}\right\} =A$, then $C$ is said to be a \textbf{%
Hamiltonian cycle }of $G$. Trivially, the complete graph has Hamiltonian
cycles.

We consider two cycles $C^{\prime }$ and $C^{\prime \prime }$ of $G,$ where
the nodes included in $C^{\prime }$ and $C^{\prime \prime }$ are represented
by $A^{\prime }$ and $A^{\prime \prime }$, respectively. We say that $%
C^{\prime }$ is \textbf{larger than} $C^{\prime \prime }$ if $A^{\prime
\prime }\subsetneq A^{\prime }$. A cycle $C$ of $G$ is said to be a \textbf{%
maximal cycle} of $G$ if there is no cycle $C^{\prime }$ of $G$ that is
larger than $C$.

We construct graphs from a given evaluability profile $\mathbf{A}$. For $%
v\in V$, let 
\begin{equation*}
G_{v}=\left\{ ab\text{ }\left\vert \text{ }a,b\in A_{v}\text{ and }a\neq
b\right. \right\} ;
\end{equation*}%
that is, if individual $v$ evaluates both $a$ and $b$, then $ab\in G_{v}$.
Moreover, 
\begin{equation*}
G=\left\{ ab\text{ }\left\vert \text{ }a,b\in A_{v}\text{ for some }v\in V%
\text{ and }a\neq b\right. \right\} .
\end{equation*}

Note that $G_{v}\subseteq G$ for all $v\in V$; that is, if there is at least
one individual who evaluates both $a$ and $b$, then $ab\in G$. If there is a
complete individual, then $G$ is the complete graph. On the other hand, in
the peer rating model, although there is no complete individual, $G$ is the
complete graph. Moreover, an evaluability profile $\mathbf{A}$ is
non-trivial if and only if $G$ is the complete graph

\section{Results}

To introduce our results, we use the following two conditions.

\begin{condition}
If there are some cycles of $G$, then for each cycle of $G$ denoted by $C$,
there is $v\in V$ such that $C\subseteq G_{v}$.
\end{condition}

Condition 1 requires that for each cycle of $G$, there is at least one
individual who evaluates all alternatives that are included in the cycle.
Note that if there is a complete individual, then Condition 1 is satisfied.
On the other hand, in the peer rating model, Condition 1 is not satisfied,
because there is a Hamiltonian cycle but no complete individual.

\begin{condition}
There is no Hamiltonian cycle of $G$\textbf{.}
\end{condition}

Condition 2 requires that $\mathbf{A}$ is sufficiently small as not to have
any Hamiltonian cycle.

We have the following result on the conditions.

\begin{lemma}
Conditions 1 is satisfied and Condition 2 is not satisfied if and only if
there is at least one complete individual.
\end{lemma}

\textbf{Proof}. First, we show the if-part. Suppose that there is at least
one complete individual denoted by $v\in V$. Then, $G$ is the complete graph
and thus has Hamiltonian cycles. Therefore, Condition 2 is not satisfied.
Moreover, each cycle of $G$ denoted by $C,$ $C\subseteq G_{v}$, because $v$
is complete individual.

Second, we show the only-if-part. Suppose that Conditions 1 is satisfied and
Condition 2 is not satisfied. Then, there is a Hamiltonian cycle. Since the
Hamiltonian cycle includes all alternatives and Condition 1 holds, there
must be at least one complete individual. \textbf{Q.E.D.}\newline

We use this result to show several results introduced later.

Now, we begin deriving the condition that ensures the existence of ARFs that
satisfy four different combinations of the properties introduced above.
First, we consider \textbf{TV} and \textbf{PC} as properties of ARFs.

We introduce an ARF $f^{\ast }$ constructed as follows. First, let $R^{\ast
}\left( \mathbf{R}\right) $ be a binary relation such that $aR^{\ast }\left( 
\mathbf{R}\right) b$ if and only if $V\left( a\right) \cap V\left( b\right)
\neq \emptyset $ and $aP\left( R_{v}\right) b$ for all $v\in V\left(
a\right) \cap V\left( b\right) $. Trivially, $R^{\ast }\left( \mathbf{R}%
\right) $ is asymmetric. Second, let 
\begin{eqnarray*}
f^{\ast }\left( \mathbf{R}\right) &\in &\mathcal{E}\left( R^{\ast }\left( 
\mathbf{R}\right) \right) \text{ if }R^{\ast }\left( \mathbf{R}\right) \text{
is acyclic,} \\
&=&\text{arbitrary weak order if otherwise.}
\end{eqnarray*}%
By Remark 1, $\mathcal{E}\left( R^{\ast }\left( \mathbf{R}\right) \right)
\neq \emptyset $ if $R^{\ast }\left( \mathbf{R}\right) $ is acyclic. Thus, $%
f^{\ast }$ is well-defined.

Then, we have the following result.

\begin{theorem}
There is an ARF that satisfies \textbf{TV} and \textbf{PC }if and only if
Condition 1 is satisfied.
\end{theorem}

\textbf{Proof. }First, we show the only-if-part. The proof of this part is
inspired by the proof of Ando et al. (2003, Theorem 3.4). Suppose that
Condition 1 is not satisfied; that is, there is a cycle of $G$ that is
defined by (\ref{c}) without $v\in V$ such that $C\subseteq G_{v}$.
Moreover, let $A^{C}=\left\{ a_{1},a_{2},\cdots ,a_{M}\right\} \subseteq A$,
which is the set of alternatives that are included in $C$.

We define a specific preference profile of individuals. First, for each $%
m=1,\cdots ,M$, let $R^{m}$ be a linear order on $A^{C}$ that satisfies 
\begin{equation*}
R^{m}:a_{m},a_{m-1},\cdots ,a_{1},a_{M},a_{M-1},\cdots ,a_{m+1}.
\end{equation*}%
Then, let $\mathbf{\bar{R}=}\left( \bar{R}_{v}\right) _{v\in V}$ be a 
\textbf{cyclic profile }of $C$\textit{\ }if for all\textit{\ }$v\in V$,%
\textit{\ }$\bar{R}_{v}$ is a linear order on $A_{v}$ satisfying 
\begin{equation*}
\left. \bar{R}_{v}\right\vert _{A^{C}}=\left. R^{m}\right\vert _{A_{v}\cap
A^{C}}
\end{equation*}%
for some $m=1,\cdots ,M$ such that $a_{m}\notin A_{v}$; that is, $a_{m}$ is
an alternative that $v$ does not evaluate.\footnote{%
Note that Ando et al. (2003) also consider a cyclic profile only in the case
where $A=V$ and $A\setminus A_{v}=\left\{ v\right\} $ for all $v\in V$.}
This profile is well-defined for any $C\subseteq G$, because there is no $%
v\in V$ such that $C\subseteq G_{v}$ due to Condition 1.

We consider characteristics of a cyclic profile of $C$ denoted by $\mathbf{%
\bar{R}}$. We arbitrary fix $m=1,\cdots ,M$. Then, $a_{m}R^{m}a_{m+1}$ but,
for each $l=1,\cdots ,M$ such that $l\neq m,$ $a_{m+1}R^{l}a_{m}$. Thus, for
each $v$ such that$\ \left. \bar{R}_{v}\right\vert _{A^{C}}=\left.
R^{m}\right\vert _{A_{v}\cap A^{C}}$, $a_{m}\notin A_{v}$. Thus, there is 
\textit{no} individual $v\in V\left( a_{m}\right) \cap V\left(
a_{m+1}\right) $ such that $a_{m}\bar{R}_{v}a_{m+1}$. Moreover, since $%
\left. \bar{R}_{v}\right\vert _{A^{C}}$ is a linear order on $A^{C}$, we
have $a_{m+1}P\left( \left. \bar{R}_{v}\right\vert _{A^{C}}\right) a_{m}$
for all $v\in V\left( a_{m}\right) \cap V\left( a_{m+1}\right) $ and all $%
m=1,\cdots ,M$. Note that since $a_{M+1}=a_{1}$, $a_{1}P\left( \left. \bar{R}%
_{v}\right\vert _{A^{C}}\right) a_{M}$ for all $v\in V\left( a_{M}\right)
\cap V\left( a_{1}\right) $.

Now, toward a contradiction, suppose that $f$ is \textbf{TV} and \textbf{PC}%
. Then, since $f$ is \textbf{PC }and the fact above holds, for all $%
m=1,\cdots ,M-1,$ $a_{m+1}P\left( f\left( \mathbf{\bar{R}}\right) \right)
a_{m}$ and $a_{1}P\left( f\left( \mathbf{\bar{R}}\right) \right) a_{M}$.
Since $f$ is \textbf{TV }and $P\left( f\left( \mathbf{\bar{R}}\right)
\right) $ is transitive, $a_{m+1}P\left( f\left( \mathbf{\bar{R}}\right)
\right) a_{m}$ for all $m=1,\cdots ,M-1$ implies $a_{M}f\left( \mathbf{\bar{R%
}}\right) a_{1}$ that contradicts $a_{1}P\left( f\left( \mathbf{\bar{R}}%
\right) \right) a_{M}$.

Second, we show the if-part. We show the following result.

\begin{claim}
If Condition 1 is satisfied, then $R^{\ast }\left( \mathbf{R}\right) $ is
acyclic for any $\mathbf{R}\in \mathcal{W}$.
\end{claim}

\textbf{Proof. }Suppose not; that is, for some $\mathbf{R}\in \mathcal{W}$, $%
R^{\ast }\left( \mathbf{R}\right) $ has a cycle for $A$; that is, there is
some $M\in \mathbb{N}\setminus \left\{ 1\right\} $ and $a_{1},\cdots ,a_{M}$
such that $a_{m+1}R^{\ast }\left( \mathbf{R}\right) a_{m}$ for all $m\in
\left\{ 1,\cdots ,M-1\right\} $ and $a_{1}R^{\ast }\left( \mathbf{R}\right)
a_{M}$. Since $R^{\ast }$ is asymmetric, $M\geq 2$ and $a_{m+1}P\left(
R^{\ast }\left( \mathbf{R}\right) \right) a_{m}$ for all $m\in \left\{
1,\cdots ,M\right\} $ where $a_{M+1}=a_{1}$. Then, since $a_{m+1}R^{\ast
}\left( \mathbf{R}\right) a_{m}$ implies $V\left( a_{m+1}\right) \cap
V\left( a_{m}\right) \neq \emptyset $, there is a cycle of $G$ denoted by (%
\ref{c}). By Condition 1, there is $v\in V$ such that $C\subseteq G_{v}$;
that is, $v\in V\left( a_{m+1}\right) \cap V\left( a_{m}\right) $ for all $%
m\in \left\{ 1,\cdots ,M\right\} $. By the definition of $R^{\ast }\left( 
\mathbf{R}\right) ,$ $a_{m+1}P\left( R_{v}\right) a_{m}$ for all $m\in
\left\{ 1,\cdots ,M\right\} $, which contradicts that $R_{v}$ is a weak
order. Therefore, $R^{\ast }$ is acyclic. \textbf{Q.E.D.}\newline

By Remark 2 and Claim 1, if Condition 1 is satisfied, then there is a linear
order extension of $R^{\ast }\left( \mathbf{R}\right) $. Therefore, in this
case, $f^{\ast }$ satisfies \textbf{TV} and \textbf{PC}. \textbf{Q.E.D.}%
\newline

Theorem 1 is a generalization of a result of Ando et al. (2003), which
discuss the peer rating model and thus have an impossibility result only.
This is because Condition 1 is not satisfied in the model and thus there is
no ARF that satisfies \textbf{TV} and \textbf{PC.}

Second, we consider an ARF that satisfies \textbf{TV}, \textbf{PC}, and 
\textbf{IIA}. We assume that Condition 1 is satisfied, because this is a
necessary condition for the existence of ARFs that satisfies \textbf{TV} and 
\textbf{PC}. We introduce the following ARF denoted by $f^{\ast \ast }$,
which is well defined under Condition 1.

We first briefly introduce points of the ARF introduced below and denoted by 
$f^{\ast \ast }$. If there are some cycles of $G$, then we choose one local
dictator of the alternatives included in each cycle. Condition 1 guarantees
that choosing a local dictator for each cycle is possible. Moreover, as
shown later, we can properly choose the local dictators to avoid any overlap.

Now, we introduce a formal description of $f^{\ast \ast }$. First, we let $%
R^{T}$ be an arbitrary linear order on $A$, which is corresponding to the
tie-breaking rule of this ARF. Second, we assume that there are some cycles
of $G$. Let$\ C^{x}$ be a maximal cycle of $G$ and $A^{x}$ be the set of
alternatives that are included in $C^{x}$. We consider the maximal cycles $%
C^{1},\cdots ,C^{X}$ satisfying

\begin{description}
\item[(i)] neither $A^{x}\setminus A^{y}$ nor $A^{y}\setminus A^{x}$ is
empty for all $x,y$ such that $x\neq y,$ and

\item[(ii)] if $a$ is not included in any of $C^{1},\cdots ,C^{X}$ ; that
is, if 
\begin{equation*}
a\in A\setminus \left( A^{1}\cup \cdots \cup A^{X}\right) \left( \equiv
A^{0}\right) \text{,}
\end{equation*}%
then $a$ is not included in any cycle of $G$.
\end{description}

By (i), there are no two of these maximal cycles $C^{1},\cdots ,C^{X}$ such
that the alternatives included in them are identical. Moreover, by (ii), any
alternative that is included in at least one cycle of $G$ is also included
in some of these maximal cycles $C^{1},\cdots ,C^{X}$.

We can construct $C^{1},\cdots ,C^{X}$ satisfying (i) and (ii) by the
following method. First, since we assume that there is one cycle, we can let 
$C^{1}$ be a maximal cycle of $G$. If $C^{1}$ satisfies (ii); that is, if $%
a\in A\setminus A^{1}$ is not included in any cycle of $G$, then $X=1;$ that
is, only one maximal cycle is formed. If otherwise, then there must exists $%
a\in A\setminus A^{1}$ that is included in a cycle of $G$. Thus, in this
case, there is a maximal cycle including $a\in A\setminus A^{1}$. Let $C^{2}$
be such a maximal cycle. Then, for $C^{1}$ and $C^{2},$ $A^{1}\setminus
A^{2}\neq \emptyset $ and $A^{2}\setminus A^{1}\neq \emptyset ,$ because $%
a\in A^{2}\setminus A^{1}$ and $C^{1}$ is a maximal cycle. Inductively, we
can construct $C^{1},\cdots ,C^{X}$ satisfying (i) and (ii).

\begin{lemma}
Suppose that there is a cycle of $G$. If Condition 1 is satisfied and $X\geq
2$, then, for all $x,y=1,\cdots ,X$ such that $x\neq y$, $C^{x}\cap
C^{y}=\emptyset $.
\end{lemma}

\textbf{Proof. }Toward a contradiction, suppose $ab\in C^{x}\cap C^{y}$.
Since there is $v\in V$ such that $C^{x}\subseteq G_{v}$ by Condition 1, for
all $c,d\in A^{x}$ such that $c\neq d$, $cd\in G$. Thus, there is a path
from $a$ to $b$ with all alternatives in $A^{x}$. Similarly, there is a path
from $b$ to $a$ with all alternatives in $A^{y}$. Therefore, there is a
cycle including all alternatives in $A^{x}\cup A^{y}$. By (i), $%
A^{x}\subsetneq A^{x}\cup A^{y}$. Therefore, this fact contradicts that $%
C^{x}$ is a maximal cycle. \textbf{Q.E.D.}\newline

Lemma 2 ensures that the ARF introduced below is well-defined.

By Condition 1, there is at least one individual $v$ such that $%
C^{x}\subseteq G_{v}$ for all $x=1,\cdots ,X$. Thus, we let $v^{x}$ be such
an individual for each $x=1,\cdots ,X$. For each $x=1,\cdots ,X$ and each $%
ab\in C^{x}$, let%
\begin{eqnarray*}
&&aP\left( R^{\ast \ast }\left( \mathbf{R}\right) \right) b\text{ if }%
aP\left( R_{v^{x}}\right) b\text{ or [}aI\left( R_{v^{x}}\right) b\text{ and 
}aR^{T}b\text{]} \\
&&bP\left( R^{\ast \ast }\left( \mathbf{R}\right) \right) a\text{ if
otherwise.}
\end{eqnarray*}%
Note that if there is no cycle of $G$, then $A^{0}=A$ and thus $R^{\ast \ast
}\left( \mathbf{R}\right) $ is empty in the construction so far.

Next, we consider each pair of alternatives $a$ and $b$ such that $a$ $\in
A^{x}$ and $b\in A^{y}$ where $x,y=0,1,\cdots ,X$ and either $x=y=0$ or $%
x\neq y$. If $V\left( a\right) \cap V\left( b\right) \neq \emptyset ,$ then
we choose $v\in V\left( a\right) \cap V\left( b\right) $ and let%
\begin{eqnarray*}
&&aP\left( R^{\ast \ast }\left( \mathbf{R}\right) \right) b\text{ if }%
aP\left( R_{v}\right) b\text{ or [}aI\left( R_{v}\right) b\text{ and }aR^{T}b%
\text{]} \\
&&bP\left( R^{\ast \ast }\left( \mathbf{R}\right) \right) a\text{ if
otherwise.}
\end{eqnarray*}%
Otherwise; that is, if $V\left( a\right) \cap V\left( b\right) =\emptyset ,$
then neither $aR^{\ast \ast }\left( \mathbf{R}\right) b$ nor $bR^{\ast \ast
}\left( \mathbf{R}\right) a$.

By (ii), if $a\neq b$ and $V\left( a\right) \cap V\left( b\right) \neq
\emptyset $, then either $aR^{\ast \ast }\left( \mathbf{R}\right) b$ or $%
bR^{\ast \ast }\left( \mathbf{R}\right) a$, and either of the two is
determined by the evaluation of one individual (and the tie-breaking rule).
Otherwise, then neither $aR^{\ast \ast }\left( \mathbf{R}\right) b$ nor $%
bR^{\ast \ast }\left( \mathbf{R}\right) a$. Therefore, $R^{\ast \ast }\left( 
\mathbf{R}\right) $ is asymmetric.

Finally, let 
\begin{eqnarray*}
f^{\ast \ast }\left( \mathbf{R}\right) &\in &\mathcal{E}\left( R^{\ast \ast
}\left( \mathbf{R}\right) \right) \text{ if }R^{\ast \ast }\left( \mathbf{R}%
\right) \text{ is acyclic} \\
&=&\text{arbitrary weak order if otherwise.}
\end{eqnarray*}

\begin{lemma}
If Condition 1 is satisfied, then $f^{\ast \ast }$ is well-defined.
\end{lemma}

\textbf{Proof.} Suppose that Condition 1 is satisfied. By Lemma 2, for each $%
(a,b)$ satisfying $a\neq b$ and $V\left( a\right) \cap V\left( b\right) \neq
\emptyset $, there is only one individual who determines either $aR^{\ast
\ast }\left( \mathbf{R}\right) b$ or $bR^{\ast \ast }\left( \mathbf{R}%
\right) a$. Moreover, by Remark 1, if $R^{\ast \ast }\left( \mathbf{R}%
\right) $ is acyclic, $\mathcal{E}\left( R^{\ast \ast }\left( \mathbf{R}%
\right) \right) \neq \emptyset $. Thus, $f^{\ast \ast }$ is well-defined. 
\textbf{Q.E.D.}\newline

We compare $f^{\ast \ast }$ with $f^{\ast }$, which is introduced earlier.
We have $R^{\ast }\left( \mathbf{R}\right) \subseteq R^{\ast \ast }\left( 
\mathbf{R}\right) $ and thus $\mathcal{E}\left( R^{\ast \ast }\left( \mathbf{%
R}\right) \right) \subseteq \mathcal{E}\left( R^{\ast }\left( \mathbf{R}%
\right) \right) $. Moreover, if $R^{\ast \ast }\left( \mathbf{R}\right) $ is
acyclic, then $R^{\ast }\left( \mathbf{R}\right) $ is also acyclic.
Therefore, if $R^{\ast \ast }\left( \mathbf{R}\right) $ is acyclic, then $%
f^{\ast \ast }\left( \mathbf{R}\right) \in \mathcal{E}\left( R^{\ast }\left( 
\mathbf{R}\right) \right) $; that is, $f^{\ast \ast }\left( \mathbf{R}%
\right) $ can be a result of $f^{\ast }\left( \mathbf{R}\right) $. This
implies that we can regard $f^{\ast \ast }\left( \mathbf{R}\right) $ as $%
f^{\ast }\left( \mathbf{R}\right) $ with a complicated tie-breaking
(incomparability-breaking) rule. We show that this rule ensures that the ARF
satisfies \textbf{IIA }under Condition 1 later.

We provide an example to understand $f^{\ast \ast }$.

\subsubsection*{Example 1}

There are seven alternatives $A=\left\{ a_{1},\ldots ,a_{7}\right\} $. There
are only three types $1,$ $2$ and $3$ of individuals with regard to their
evaluable set. Suppose that $v_{i}$ is a type $i=1,2,3$ individual. The
evaluable sets of $1,2$ and $3$ are given by 
\begin{eqnarray*}
A_{v_{1}} &=&\{a_{1},a_{2},a_{3},a_{4}\}, \\
A_{v_{2}} &=&\{a_{4},a_{5},a_{6}\}, \\
A_{v_{3}} &=&\{a_{6},a_{7}\}.
\end{eqnarray*}%
Moreover, the evaluable set of any other individual is either of the three.
Then, $G$ is written in Figure 1.

\begin{figure}[tbph]
\begin{center}
\includegraphics[width=75mm]{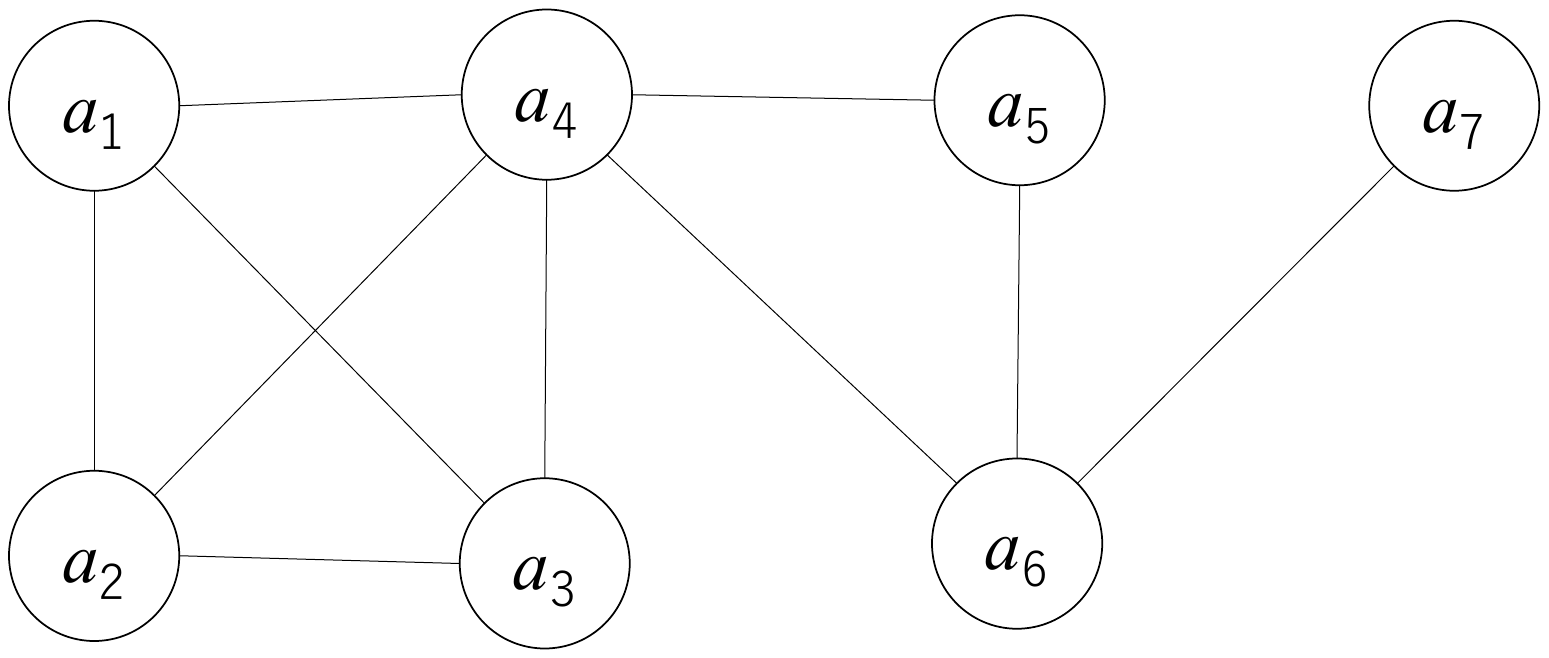}
\end{center}
\caption{$G$ of Example 1}
\end{figure}

Then, there are two maximal cycles%
\begin{eqnarray*}
C^{1} &=&\{a_{1}a_{2},a_{2}a_{3},a_{3}a_{4},a_{4}a_{1}\}, \\
C^{2} &=&\{a_{4}a_{5},a_{5}a_{6},a_{6}a_{4}\},
\end{eqnarray*}%
and $A^{0}=\left\{ a_{7}\right\} $. We let $v^{1}=v_{1}$ and $v^{2}=v_{2};$
that is, relative positions of $a_{1},\cdots ,a_{4}$ and those of $%
a_{4},a_{5},a_{6}$ are determined by $v_{1}$ and $v_{2},$ respectively.
Moreover, we assume that relative positions of $a_{6}$ and $a_{7}$ is
determined by $v_{3}$.

Now, let 
\begin{eqnarray*}
R_{v_{1}} &:&a_{2},a_{4},a_{1},a_{3} \\
R_{v_{2}} &:&a_{5},a_{6},a_{4}, \\
R_{v_{3}} &:&a_{7},a_{6}
\end{eqnarray*}%
Then, 
\begin{eqnarray*}
R^{\ast \ast }\left( \mathbf{R}\right) &=&R_{v_{1}}\cup R_{v_{2}}\cup
R_{v_{3}} \\
&=&\{\left( a_{2},a_{4}\right) ,\left( a_{2},a_{1}\right) ,\left(
a_{2},a_{3}\right) ,\left( a_{4},a_{1}\right) ,\left( a_{4},a_{3}\right) , \\
&&\left( a_{1},a_{3}\right) ,\left( a_{5},a_{6}\right) ,\left(
a_{5},a_{4}\right) ,\left( a_{6},a_{4}\right) ,\left( a_{7},a_{6}\right) \},
\end{eqnarray*}%
which is acyclic. Then, since the outcome of this function is a linear order
extension of $R^{\ast \ast }\left( \mathbf{R}\right) $, for example, 
\begin{equation*}
f^{\ast \ast }\left( \mathbf{R}\right)
:a_{2},a_{5},a_{7},a_{6},a_{4},a_{1},a_{3}\text{.}
\end{equation*}%
\newline

We introduce the following result.

\begin{theorem}
There is an ARF that satisfies \textbf{TV}, \textbf{PC}, and \textbf{IIA }if
and only if Condition 1 is satisfied.
\end{theorem}

\textbf{Proof. }By Theorem 1, Condition 1 is trivially necessary. Thus, we
show the if-part only. We show that if Condition 1 is satisfied, then $%
f^{\ast \ast }$ satisfies \textbf{TV}, \textbf{PC}, and \textbf{IIA}.

First, we show the following result.

\begin{claim}
If Condition 1 is satisfied, then $R^{\ast \ast }\left( \mathbf{R}\right) $
is acyclic for any $\mathbf{R}\in \mathcal{W}$.
\end{claim}

\textbf{Proof.} Suppose not; that is, there are distinct $a_{1},a_{2},\cdots
,a_{M}\in A$ such that $a_{m+1}R^{\ast \ast }\left( \mathbf{R}\right) a_{m}$
for all $m=1,\cdots ,M$. Then, there exists a cycle that is defined by (\ref%
{c}) and thus there also exists $C^{x}$ such that $\left\{
a_{1},a_{2},\cdots ,a_{M}\right\} \subseteq A^{x}$. Then, for all $%
m=1,\cdots ,M$, either $a_{m+1}P\left( R_{v^{x}}\right) a_{m}$ or [$%
a_{m+1}I\left( R_{v^{x}}\right) a_{m}$ and $a_{m+1}R^{T}a_{m}$]. Therefore,
in any case, $a_{m+1}R_{v^{x}}a_{m}$ for all $m=1,\cdots ,M$. By the
transitivity of $R_{v^{x}}$ and $a_{M+1}=a_{1}$, $a_{m}R_{v^{x}}a_{m+1}$ for
all $m=1,\cdots ,M$. Hence, $a_{m+1}I\left( R_{v^{x}}\right) a_{m}$ and thus 
$a_{m+1}R^{T}a_{m}$ for all $m=1,\cdots ,M$, but it contradicts the
transitivity of $R^{T}$. Therefore, $R^{\ast \ast }\left( \mathbf{R}\right) $
is acyclic. \textbf{Q.E.D.}\newline

Second, we show that if $R^{\ast \ast }\left( \mathbf{R}\right) $ is
acyclic, $f^{\ast \ast }$ satisfies \textbf{IIA}. Let $a,b\in A$ such that $%
a\neq b$ and $V\left( a\right) \cap V\left( b\right) \neq \emptyset $, $%
\mathbf{R}$ and $\mathbf{R}^{\prime }$ such that $\left. R_{v}\right\vert
_{\{a,b\}}=\left. R_{v}^{\prime }\right\vert _{\{a,b\}}$ for all $v\in
V\left( a\right) \cap V\left( b\right) $. Then, by the construction, there
is $v^{\ast }\in V\left( a\right) \cap V\left( b\right) $ such that 
\begin{eqnarray*}
&&aP\left( f^{\ast \ast }\left( \mathbf{R}\right) \right) b\text{ if }%
aP\left( R_{v^{\ast }}\right) b\text{ or [}aI\left( R_{v^{\ast }}\right) b%
\text{ and }aR^{T}b\text{]} \\
&&bP\left( f^{\ast \ast }\left( \mathbf{R}\right) \right) a\text{ if
otherwise.}
\end{eqnarray*}%
Since $\left. R_{v^{\ast }}\right\vert _{\{a,b\}}=\left. R_{v^{\ast
}}^{\prime }\right\vert _{\{a,b\}}$, $\left. f^{\ast \ast }\left( \mathbf{R}%
\right) \right\vert _{\{a,b\}}=\left. f^{\ast \ast }\left( \mathbf{R}%
^{\prime }\right) \right\vert _{\{a,b\}}$. Therefore, $f^{\ast \ast }$
satisfies \textbf{IIA}.

Finally, we show that if $R^{\ast \ast }\left( \mathbf{R}\right) $ is
acyclic, then $f^{\ast \ast }$ satisfies \textbf{PC}. If $aP\left(
R_{v}\right) b$ for all $v\in V\left( a\right) \cap V\left( b\right) $, then 
$aP\left( R^{\ast \ast }\left( \mathbf{R}\right) \right) b$. Therefore, $%
f^{\ast \ast }$ satisfies \textbf{PC}. Then, by Claim 2, if Condition 1 is
satisfied, then $f^{\ast \ast }$ satisfies \textbf{TV}, \textbf{PC}, and 
\textbf{IIA}. \textbf{Q.E.D.}\newline

Condition 1 guarantees the existence of an ARF that satisfies the three
properties. Thus, the if-part of Theorem 1 is generalized. On the other
hand, the only-if-part of Theorem 2 is implied by that of Theorem 1.

We consider some sufficient conditions for the existence of an ARF $f$ that
satisfies the three properties in order to clarify the result.

\begin{corollary}
If $G$ has no cycle, then there is an ARF that satisfies \textbf{TV}, 
\textbf{PC}, and \textbf{IIA}.
\end{corollary}

This result implies that if $\mathbf{A}$ is sufficiently small, then there
exists an ARF that satisfies the three properties. We briefly show this
fact. Let $\mathbf{A}^{\prime }$ be a smaller evaluability profile than $%
\mathbf{A}$ and $G^{\prime }$ be the graph constructed from the problem $%
(V,A,\mathbf{A}^{\prime })$. Then, $G^{\prime }\subseteq G$. Thus, if $G$
has no cycle, then $G^{\prime }$ also has no cycle.

\begin{corollary}
If there is at least one complete individual, then there is an ARF that
satisfies \textbf{TV}, \textbf{PC}, and \textbf{IIA}.
\end{corollary}

This result is trivial from Lemma 1 and Theorem 2. This implies that if an
evaluability profile $\mathbf{A}$ is sufficiently large, then there is an
ARF $f$ that satisfies the three properties. This is because, if there is a
complete individual\textbf{, }then the individual must also be complete for
a larger evaluability profile.

By Corollaries 1 and 2, if $\mathbf{A}$ is either sufficiently small or
sufficiently large, then there is an ARF that satisfies the three
properties. Conversely, if $\mathbf{A}$ is neither very small nor very
large, then an ARF $f$ that satisfies \textbf{TV} and \textbf{PC }may fail
to exist.

Next, we focus on the situation where Condition 2 is not satisfied.

\begin{corollary}
Suppose that Condition 2 is not satisfied. Then, there is an ARF that
satisfies \textbf{TV}, \textbf{PC}, and \textbf{IIA }if and only if there is
at least one complete individual. Moreover, this is satisfied even without 
\textbf{IIA}.
\end{corollary}

This result is straightforward from Lemma 1, Theorems 1 and 2. Note that in
the peer rating model, there is a Hamiltonian cycle and there is no complete
individual. Thus, as also shown by Ando et al. (2003), there is no ARF that
satisfies \textbf{TV} and \textbf{PC }in the model.

Third, we additionally require \textbf{ND}. We first consider the case where
at least one individual is complete. By Corollary 2, in this case, there is
an ARF that satisfies \textbf{TV}, \textbf{PC}, and \textbf{IIA}. Ohbo et
al. (2005) show the following result.

\begin{remark}
Suppose that there is at least one complete individual. If an ARF $f$
satisfies \textbf{TV}, \textbf{PC}, and \textbf{IIA}, then there is a
dictator of $f$.
\end{remark}

Since all individuals are complete\textbf{\ }in the model of Arrow (1963),
Remark 3 is a generalization of Arrow's impossibility result. Here, we
further generalize Remark 3.

\begin{theorem}
There is an ARF that satisfies \textbf{TV}, \textbf{PC}, \textbf{IIA}, and%
\textbf{\ ND }if and only if Conditions 1 and 2 are satisfied.
\end{theorem}

\textbf{Proof.} We show the only-if-part. By Remark 4, it is necessary that
no complete individual exists. By Corollary 3, if there is no complete
individual but $G$ has a Hamiltonian cycle, then there is no ARF $f$
satisfies \textbf{TV} and \textbf{PC}. Thus, Condition 2 is a necessary
condition. Moreover, by Theorem 2, Condition 1 is also necessary.

Next, show the if-part. Since there is no Hamiltonian cycle, any individual%
\textbf{\ }$v$ is not complete. Hence $f^{\ast \ast }$ must satisfies 
\textbf{ND}. Thus, by this fact and Theorem 2, we have the if-part of
Theorem 3. \textbf{Q.E.D.}\newline

By this result, if we require \textbf{ND }in addition to others, then
Condition 2 is a necessary (and sufficient) additional condition to ensure
the existence of an ARF satisfying all of them. By Corollary 2, there is an
ARF satisfying them excepting \textbf{ND }if $\mathbf{A}$ is sufficiently
large. However, if \textbf{ND }is additionally required, then there is no
ARF satisfying them if $\mathbf{A}$ is sufficiently large. In other words,
when an evaluability profile $\mathbf{A}$ is large as to have a Hamiltonian
cycle, an ARF that satisfies \textbf{TV}, \textbf{PC}, and \textbf{IIA},
must be a dictatorship rule.

On the other hand, we immediately have the following result.

\begin{corollary}
If $G$ has no cycle, then there is an ARF that satisfies \textbf{TV}, 
\textbf{PC}, \textbf{IIA}, and\textbf{\ ND}.
\end{corollary}

This result is stronger than Corollary 1 and implies that if $\mathbf{A}$ is
sufficiently small, then there is an ARF other than the dictatorship ones
that satisfies \textbf{TV}, \textbf{PC}, and \textbf{IIA}.

Next, we focus on the class of problems where the evaluability profiles are
non-trivial. This class is discussed by Fujiwara-Greve et al. (2023). In
this class, we have the following impossibility result.

\begin{corollary}
If $A$ is non-trivial, then there is no ARF that satisfies \textbf{TV}, 
\textbf{PC}, \textbf{IIA}, and\textbf{\ ND}.
\end{corollary}

As noted above, $A$ is non-trivial if and only if $G$ is complete. First, if
there is no complete individual, then Condition 1 is not satisfied. By
Theorem 2, in this case, there is no ARF that satisfies \textbf{TV}, \textbf{%
PC}, and \textbf{IIA}. Second, if there is a complete individual, then, by
Remark 4, any ARF $f$ satisfying \textbf{TV}, \textbf{PC}, and \textbf{IIA}
does not satisfy \textbf{ND}. Therefore, we have Corollary 5.\newline

Next, we consider a \textbf{wPC} and \textbf{NC }instead of \textbf{PC},
because they are weaker than \textbf{PC} because of Remark 1. First, we have
the following result.

\begin{lemma}
Suppose that an ARF $f$ satisfies \textbf{TV}, \textbf{wPC},\ and\textbf{\
IIA}, and Condition 1 is not satisfied; that is, there is a cycle of $G$
that is defined by (\ref{c}) without $v\in V$ such that $C\subseteq G_{v}$.
For any $\mathbf{R}\in \mathcal{W},$ and any $l,n\in \left\{ 1,\cdots
,M\right\} $ such that $l\neq n$, $a_{l}I\left( f\left( \mathbf{R}\right)
\right) a_{n}$.
\end{lemma}

\textbf{Proof.} This proof is inspired by that of Ando et al. (2003, Theorem
3.7). First, we show the following result.

\begin{claim}
Suppose that an ARF $f$ satisfies \textbf{TV}, \textbf{wPC}, and\textbf{\ IIA%
}, and Condition 1 is not satisfied; that is, there is a cycle of $G$ that
is defined by (\ref{c}) without $v\in V$ such that $C\subseteq G_{v}$. For
any $n=1,\cdots ,M$, if $a_{n+1}P\left( R_{v}\right) a_{n}$ for all $v\in
V\left( a_{n}\right) \cap V\left( a_{n+1}\right) $, then $a_{n+1}I\left(
f\left( \mathbf{R}\right) \right) a_{n}$, where $a_{M+1}=a_{1}$.
\end{claim}

\textbf{Proof.} We can let $\mathbf{\bar{R}=}\left( \bar{R}_{v}^{C}\right)
_{v\in V}$ be a cyclic profile\textbf{\ }of $C$ that is defined in the proof
of Theorem 1, because there is no $v\in V$ such that $C\subseteq G_{v}$.
Then, for all $m=1,\cdots ,M$, $a_{m+1}P\left( \bar{R}_{v}\right) a_{m}$ for
all $v\in V\left( a_{m}\right) \cap V\left( a_{m+1}\right) $. Since $f$
satisfies \textbf{wPC}, for any $n=1,\cdots ,M,$ $a_{n+1}f\left( \mathbf{%
\bar{R}}\right) a_{n}$. Then, by the transitivity of $f\left( \mathbf{\bar{R}%
}\right) $ and $a_{M+1}=a_{1}$, we also have $a_{n}f\left( \mathbf{\bar{R}}%
\right) a_{n+1}$ and thus $a_{n+1}I\left( f\left( \mathbf{\bar{R}}\right)
\right) a_{n}$ for any $n=1,\cdots ,M$.

Now, we consider an arbitrary $\mathbf{R}$ such that $a_{n+1}P\left(
R_{v}\right) a_{n}$ for all $v\in V\left( a_{n}\right) \cap V\left(
a_{n+1}\right) $. Then, since 
\begin{equation*}
\left. R_{v}\right\vert _{\{a_{n},a_{n+1}\}}=\left. \bar{R}_{v}\right\vert
_{\{a_{n},a_{n+1}\}}
\end{equation*}%
for all $v\in V$ and $f$ satisfies\textbf{\ IIA}, we have $a_{n+1}I\left(
R_{v}\right) a_{n}$. \textbf{Q.E.D.}\linebreak

Next, we show that for any $\mathbf{R}\in W_{V},$ and for any $n=1,\cdots ,M$%
, $a_{n+1}I\left( f\left( \mathbf{R}\right) \right) a_{n}$, where $%
a_{M+1}=a_{1}$ and $A^{C}=\left\{ a_{1},\cdots ,a_{M}\right\} $ be the set
of alternatives that are included in cycle $C$ defined in Lemma 4.

Without loss of generality, we let $n=1$ and show $a_{2}I\left( f\left( 
\mathbf{R}\right) \right) a_{1}$. Let $\mathbf{R}$ be an arbitrary
preference profile of individuals. We introduce another preference profile
of individuals denoted by $\mathbf{R}^{\prime }$ that have the following
characteristics. First, let $R^{\ast }$ be a linear order on $A^{C}$
satisfying 
\begin{equation*}
R^{\ast }:a_{M},a_{M-1},\cdots ,a_{2},a_{1}.
\end{equation*}%
Then, let $\mathbf{R}^{\prime }$ be such that (i) $\left. R_{v}^{\prime
}\right\vert _{\{a_{1},a_{2}\}}=\left. R_{v}\right\vert _{\{a_{1},a_{2}\}}$
for all $v\in V\left( a_{n}\right) \cap V\left( a_{n+1}\right) $, (ii) for
all $m\in \{2,\cdots ,M\}$%
\begin{equation*}
\left. R_{v}^{\prime }\right\vert _{\{a_{m},a_{m+1}\}}=\left. R^{\ast
}\right\vert _{\{a_{m},a_{m+1}\}}
\end{equation*}%
for all $v\in V\left( a_{m}\right) \cap V\left( a_{m+1}\right) $. Since $%
a_{l}P\left( R^{\ast }\right) a_{2}$ and $a_{l}P\left( R^{\ast }\right)
a_{1} $ for all $l\in \{3,\cdots ,M\}$, $a_{2}R_{v}^{\prime }a_{1}$ or $%
a_{1}R_{v}^{\prime }a_{2}\,$(or both) is possible. Therefore, $\mathbf{R}%
^{\prime }$ is well-defined.

By Claim 3, $a_{m+1}I\left( f\left( \mathbf{R}^{\prime }\right) \right)
a_{m} $ for all $m\in \{2,\cdots ,M\}$. Therefore, by the transitivity of $%
I\left( f\left( \mathbf{R}^{\prime }\right) \right) $, $a_{2}I\left( f\left( 
\mathbf{R}^{\prime }\right) \right) a_{1}$. Moreover, since $\left.
R_{v}^{\prime }\right\vert _{\{a_{1},a_{2}\}}=\left. R_{v}\right\vert
_{\{a_{1},a_{2}\}}$ for all $v\in V\left( a_{m}\right) \cap V\left(
a_{m+1}\right) $ and $f$ satisfies \textbf{IIA}, we have $a_{2}I\left(
f\left( \mathbf{R}\right) \right) a_{1}$.

Then, for any $\mathbf{R}\in W_{V},$ and for any $n=1,\cdots ,M$, $%
a_{n+1}I\left( f\left( \mathbf{R}\right) \right) a_{n}$, where $%
a_{M+1}=a_{1} $. By the transitivity of $I\left( f\left( \mathbf{R}\right)
\right) $, for any $l,n\in \left\{ 1,\cdots ,M\right\} $ such that $l\neq n$%
, $a_{l}I\left( f\left( \mathbf{R}\right) \right) a_{n}$. \textbf{Q.E.D.}%
\newline

By Lemma 4, we have the following result.

\begin{theorem}
There is an ARF $f$ satisfies \textbf{TV}, \textbf{wPC},\textbf{\ NC}, and\ 
\textbf{IIA }if and only if Condition 1 is satisfied.
\end{theorem}

\textbf{Proof.} By Theorem 2, the only-if-part is trivial.

We show the if-part. By Lemma 4, if Condition 1 is not satisfied; that is,
if there is a cycle of $G$ that is defined by (\ref{c}) where there is no $%
v\in V$ such that $C\subseteq G_{v}$, then any two alternatives included in
the cycle are always indifferent with any ARF satisfying \textbf{TV}, 
\textbf{wPC}, and\ \textbf{IIA}. Therefore, in this case, any ARF satisfying 
\textbf{TV}, \textbf{wPC}, and\ \textbf{IIA }does not satisfy \textbf{NC}. 
\textbf{Q.E.D.}\newline

Thus, even if we require \textbf{wPC }and \textbf{NC }instead of \textbf{PC}%
, Condition 1 is still necessary. Ando et al. (2003) show that in the peer
rating model, there is no ARF $f$ that satisfies \textbf{TV}, \textbf{NC},
and\ \textbf{IIA}. The necessary and sufficient condition that there is an
ARF $f$ that satisfies \textbf{TV}, \textbf{NC}, and\ \textbf{IIA }is
unknown. Ohseto (2007) also examines the peer rating model and shows that
the Borda rule satisfies \textbf{TV}, \textbf{wPC}, and\textbf{\ NC}, but it
does not satisfy \textbf{IIA}.\footnote{%
Ohseto (2007) does not mention that the rule satisfies \textbf{NC}, but it
is trivially satisifed.}

\section{Proportions of evaluability profiles}

By Theorems 2 and 3, we can classify the evaluability profile into three
types. First one is that for which there is no ARF that satisfies \textbf{TV}%
, \textbf{PA}, and\ \textbf{IIA}; that is, neither Condition 1 nor 2 is
satisfied with this type of profile. Hereafter, we refer this type of
evaluability profile as an \textquotedblleft \textbf{impossible%
\textquotedblright\ profile (}hereafter \textit{IP}\textbf{)}. Second one is
that for which an ARF that satisfies \textbf{TV}, \textbf{PA}, and\ \textbf{%
IIA }must be a dictatorship rule; that is, Condition 1 is satisfied but
Condition 2 is not with this type of profile. Hereafter, we refer this type
of evaluability profile as a \textbf{dictatorship profile (}hereafter 
\textit{DP}\textbf{)}. Third one is that for which there is an ARF that
satisfies \textbf{TV}, \textbf{PA}, \textbf{IIA}, and\textbf{\ ND}; that is,
Conditions 1 and 2 are satisfied with this type of profile. To discuss how
frequently each type occurs, we derive the proportion of each type.
Hereafter, we refer this type of evaluability profile as a \textquotedblleft 
\textbf{possible\textquotedblright\ profile (}hereafter \textit{PP}\textbf{)}%
. Note that an \textit{IP} is such that \textbf{TV}, \textbf{PA}, and\ 
\textbf{IIA} are impossible, and a \textit{DP} is such that \textbf{ND }as
well as the three are possible.

We derive all possible evaluable sets of an individual. For example, suppose 
$A=\left\{ a,b,c\right\} $. Then, the potential evaluable sets for an
individual are $\left\{ a,b,c\right\} $, $\left\{ a,b\right\} $, $\left\{
a,c\right\} $, and $\left\{ b,c\right\} $, because each of them must include
at least two alternatives. In general, there are $2^{\left\vert A\right\vert
}-\left\vert A\right\vert -1$ potential evaluable sets for each individual.
Therefore, in total, there are $\left( 2^{\left\vert A\right\vert
}-\left\vert A\right\vert -1\right) ^{\left\vert I\right\vert }$ potential
evaluability profiles.

First, we consider the proportion of the \textit{DP}s. By Lemma 1, in this
type, there is at least one complete individual for the evaluability
profile. Therefore, there are \textit{DP}s that account for 
\begin{equation}
1-\left( \frac{2^{\left\vert A\right\vert }-\left\vert A\right\vert -2}{%
2^{\left\vert A\right\vert }-\left\vert A\right\vert -1}\right) ^{\left\vert
I\right\vert }  \label{a}
\end{equation}%
of the total, and the remainder consists of either \textit{IP}s or \textit{PP%
}s. In each cell of the following table, the approximation of (\ref{a}) with
the corresponding combination of $\left\vert I\right\vert $ and $\left\vert
A\right\vert $ is written. By the table, if $\left\vert A\right\vert $ is
small, then in many evaluability profiles, there is at least one complete
individual and the ARF satisfying \textbf{TV}, \textbf{PA}, and\ \textbf{IIA 
}must be a dictatorship rule. On the other hand, if $\left\vert A\right\vert 
$ is large, then almost evaluability profiles is such that there is no ARF
satisfying \textbf{TV}, \textbf{PA}, and\ \textbf{IIA }or there is some ARF
satisfying \textbf{TV}, \textbf{PA},\ \textbf{IIA},\textbf{\ }and\textbf{\
ND.} \newline

\begin{center}
\begin{tabular}{|c|c|c|c|c|c|c|c|c|c|}
\hline
$\left\vert A\right\vert \setminus \left\vert I\right\vert $ & 3 & 6 & 9 & 12
& 15 & 18 & 21 & 24 & 27 \\ \hline
3 & 0.58 & 0.82 & 0.92 & 0.97 & 0.99 & 0.99 & 1 & 1 & 1 \\ \hline
5 & 0.11 & 0.21 & 0.30 & 0.38 & 0.44 & 0.51 & 0.56 & 0.61 & 0.65 \\ \hline
7 & 0.02 & 0.05 & 0.07 & 0.10 & 0.12 & 0.14 & 0.16 & 0.18 & 0.20 \\ \hline
9 & 0.00 & 0.01 & 0.02 & 0.02 & 0.03 & 0.04 & 0.04 & 0.05 & 0.05 \\ \hline
\end{tabular}

\textbf{Table 1: }Proportions of \textit{DP}s \\[0pt]
\end{center}

Deriving the proportion of \textit{IP}s and that of \textit{PP}s is
relatively difficult, even if neither $\left\vert A\right\vert $ nor $%
\left\vert I\right\vert $ is very large. Here, we derive them only the three
cases where $\left( \left\vert I\right\vert ,\left\vert A\right\vert \right)
=\left( 3,3\right) ,$ $\left( 4,3\right) $ and $\left( 3,4\right) $.

First, if $\left( \left\vert I\right\vert ,\left\vert A\right\vert \right)
=\left( 3,3\right) ,$ then there are $6$, $37$, and $21$ evaluability
profiles that are \textit{IP}s, \textit{DP}s, and \textit{PP}s,
respectively. Thus, about $33\%$ of all profiles are for which there is an
ARF that satisfies \textbf{TV}, \textbf{PA},\ \textbf{IIA},\textbf{\ }and%
\textbf{\ ND}. Second, if $\left( \left\vert I\right\vert ,\left\vert
A\right\vert \right) =\left( 4,3\right) ,$ then there are $36$, $175$, and $%
45$ evaluability profiles that are \textit{IP}s, \textit{DP}s, and \textit{PP%
}s, respectively. Thus, about $18\%$ of all profiles are for which there is
an ARF that satisfies the four properties. By comparing the two cases, the
proportion of \textit{PP}s may decrease with the number of individuals.
Third, if $\left( \left\vert I\right\vert ,\left\vert A\right\vert \right)
=\left( 3,4\right) ,$ then there are $371$, $331$, and $629$ evaluability
profiles that are \textit{IP}s, \textit{DP}s, and \textit{PP}s,
respectively. Thus, about $47\%$ of all profiles are for which there is an
ARF that satisfies the four properties. By comparing the cases where $\left(
\left\vert I\right\vert ,\left\vert A\right\vert \right) =\left( 3,3\right) $
and $\left( 3,4\right) $, the proportion of \textit{PP}s may increase with
the number of individuals.

While this analysis provides only limited conclusions, it suggests that when
the number of individuals is much larger than that of alternatives,
requiring \textbf{TV}, \textbf{PA}, \textbf{IIA}, and\textbf{\ ND }may be
almost impossible, whereas when the number of alternatives is much larger
than that of individuals, it may be almost possible.

\section{Concluding Remarks}

In the social welfare function \`{a} la Arrow, it is assumed that each
individual evaluates all alternatives. Moreover, in real voting systems such
as national elections, we may be supposed to vote after evaluating all
alternatives. However, we are sometimes not well acquainted enough to
evaluate all alternatives.

Here, we consider a situation where some individuals lack information about
a certain alternative, while other individuals are well-informed about and
prefer that alternative. Moreover, the other alternatives are known by all
individuals. In such a case, since individuals are likely to be risk-averse,
many of those who are not well-informed about the alternative may express
their preferences by ranking it lower than other alternatives. Consequently,
that alternative would end up with a low position in the aggregated ranking.
On the other hand, as discussed in this study, if no individual evaluates
their unknown alternatives, then the alternative may be ranked higher in the
aggregated ranking.

Therefore, although whether or not the rule \textquotedblleft no individual
evaluates their unknown alternatives\textquotedblright\ should be adopted is
debatable, the decision to adopt this rule may have a significant impact on
the aggregated ranking.

\section*{Reference}

\begin{description}
\item Adachi, T., 2014. A natural mechanism for eliciting ranking when
jurors have favorites. Games and Economic Behavior 87, 508--518.

\item Amor\'{o}s, P., 2009. Eliciting socially optimal rankings from unfair
jurors. Journal of Economic Theory 144, 1211--1226.

\item Amor\'{o}s, P., 2011. A natural mechanism to choose the deserving
winner when the jury is made up of all contestants. Economics Letters 110,
241--244.

\item Amor\'{o}s, P., 2019. Choosing the winner of a competition using
natural mechanisms: Conditions based on the jury, Mathematical Social
Sciences 98 26--38

\item Amor\'{o}s, P., Corch\'{o}n, L., Moreno, B., 2002. The scholarship
assignment problem. Games and Economic Behavior 38, 1--18.

\item Ando, K., Ohara, A., Yamamoto, Y. 2003. Impossibility theorems on
mutual evaluation (in Japanese), Journal of the Operations Research Society
of Japan 46, 523--532.

\item Ando, K., Tsurutani, M. Umezawa, M., Yamamoto, Y. 2007. Impossibility
and possibility theorems for social choice functions on incomplete
preference profiles, Pacific Journal of Optimization 3(1), 11-25.

\item Ariely, D. 2010. The Upside of Irrationality: The Unexpected Benefits
of Defying Logic at Work and at Home, HarperCollins, New York.

\item Arrow, K.J., 1963. Social Choice and Individual Values, 2nd Edition.
Wiley, New York.

\item Barber\'{a}, S., Bossert W. 2023. Opinion aggregation: Borda and
Condorcet revisited, Journal of Economic Theory 210, 105654.

\item Cengelci, M. A., Sanver, M. R. 2007. Is abstention an escape from
Arrow's theorem? Social Choice and Welfare 28, 439--442.

\item Duggan, J. 1999. A General Extension Theorem for Binary Relations,
Journal of Economic Theory 86, 1-16.

\item Edelman, P.H., Por, A., 2021. A new axiomatic approach to the
impartial nomination problem. Games and Economic Behavior 130, 443--451.

\item Fujiwara-Greve, T., Kawada, Y., Nakamura, Y., Okamoto, N., 2023.
Accountable Voting. Available at SSRN: https://ssrn.com/abstract=4406802

\item Hansson, B. 1973. The independence condition in the theory of social
choice. Theory and Decision 4, 25--49.

\item Holzman, R., Moulin, H., 2013. Impartial nominations for a prize.
Econometrica 81, 173-196.

\item Kitahara, M., Okumura, Y. 2023. On Extensions of Partial Priorities in
School Choice. Mimeo Available at SSRN: https://ssrn.com/abstract=4462665

\item Mackenzie, A., 2015. Symmetry and impartial lotteries. Games and
Economic Behavior 94, 15--28.

\item Mackenzie, A., 2019. Impartial nominations for a prize. Economic
Theory 69, 713--743.

\item Ng, Y.K., Sun, G.Z. 2003. Exclusion of self evaluations in peer
ratings: an impossibility and some proposals, Social Choice and Welfare 20,
443--456.

\item Ohseto, S. 2007. A characterization of the Borda rule in peer ratings,
Mathematical Social Sciences 54(2) 147-151.

\item Ohbo, K., Tsurutani, M., Umezawa, M., Yamamoto, Y. 2005. Social
welfare function for restricted individual preference domain, Pacific
Journal of Optimization 1(2), 315-325.

\item Quesada, A. 2005. Abstention as an escape from Arrow's theorem. Social
Choice and Welfare, 25 221--226.

\item Tamura, S. 2016. Characterizing minimal impartial rules for awarding
prizes, Games and Economic Behavior 95, 41-46.

\item Tamura, S., Ohseto, S., 2014. Impartial nomination correspondences,
Social Choice and Welfare 43, 47-54.

\item Villar, A., 2023. The precedence function: a numerical evaluation
method for multi criteria ranking problems, Economic Theory Bulletin 11,
211--219.
\end{description}

\end{document}